\documentstyle[twocolumn,aps,floats]{revtex}

\begin{document}

\draft \tolerance = 10000

\setcounter{topnumber}{1}
\renewcommand{\topfraction}{0.9}
\renewcommand{\textfraction}{0.1}
\renewcommand{\floatpagefraction}{0.9}

%Fixing abstract in twocolumn mode
\twocolumn[\hsize\textwidth\columnwidth\hsize\csname
@twocolumnfalse\endcsname

\title{Can a Particle's Velocity Exceed the Speed of Light in Empty Space?}
\author{L.Ya.Kobelev   \\
Department of  Physics, Urals State University \\
 Av. Lenina, 51, Ekaterinburg 620083, Russia \\
 E-mail:  leonid.kobelev@usu.ru  }
\maketitle

\begin{abstract}
Relative motion in space with multifractal time (fractional dimension of
time close to integer $d_{t}=1+\varepsilon ({\mathbf r},t),\,|\varepsilon
|\ll 1$) for "almost" inertial frames of reference (time is almost
homogeneous and almost isotropic) is considered. Presence in such space of
absolute frames of reference and violation of conservation laws (though,
small because of the smallness of $\varepsilon$) due to the openness of
all physical systems and inhomogeneiy of time are shown. The total energy
of a body moving with $v=c$ is obtained to be finite and modified Lorentz
transformations are formulated. The relation for the total energy (and the
whole theory) reduce to the known formula of the special relativity in
case of transition to the usual time with dimension equal to unity.
\end{abstract}

\vspace{1cm}

%Fixing abstract in twocolumn mode
]

\section{Introduction}

As well known, the special relativity theory (SR) is the theory of
inertial systems and for such systems the answer to the question posed in
the title of this paper is negative. But in the nature ideal inertial
systems do not exist. It allows to raise the following problem: is it
possible to develop a theory of systems close to inertial ("almost"
inertial systems) that would include, as a special case, special
relativity but, at the same time, would allow for a motion of particles
with any velocity? Obviously, in order to invent such a theory it is
necessary to refuse from  the rigorous validity of any of the SR
postulates: the homogenity of space and time, the invariance of speed of
light and the Galilee invariance principle.

The present paper suggests an example of such a theory, based on the
concepts of time and space with fractional dimensions (FD) developed in
the theory of multifractal time and space \cite{kob1}. We begin with the
first of the mentioned SR principles. In an inhomogeneous space and time,
if the inhomogenities are small enough, any motion will be close to that
in homogeneous space ("almost" inertial), but the velocity of light can
alter slightly, being thus "almost" constant. Mere assumption that the
values of fractional dimensions of space and time are close to integer
leads then to conclusion that the motions of particles with any velocities
become possible. Other main assumptions made are the light velocity
invariance and invariance with respect to modified Lorentz
transformations. The results our theory gives for the velocities less than
the velocity of light $c$ almost coincide with SR, but it does not contain
singularities at $v=c$. Stress, that this theory is not a generalization
of SR theory, because any such generalization in the domain of SR validity
(inertial systems) is absurd. Our theory describes relative movements only
in the "almost" inertial systems, and thus does not contradict to SR.

\section{Multifractal time}

Following \cite{kob1}, we will consider both time and space as the only
material fields existing in the world and generating all other physical
fields. Assume that every of them consists of a continuous, but not
differentiable bounded set of small elements (elementary intervals,
further treated as "points"). Consider the set of small time elements
$S_{t}$. Let time be defined on multifractal subsets of such elements,
defined on certain measure carrier ${\mathcal R}^{n}$. Each element of
these subsets (or "points") is characterized by the local fractional
(fractal) dimension (FD) $d_{t}({\mathbf r}(t),t)$ and for different
elements FD are different. In this case the classical mathematical
calculus or fractional (say, Riemann - Liouville) calculus \cite{sam} can
not be applied to describe a small changes of a continuous function of
physical values $f(t)$, defined on time subsets $S_{t}$, because the
fractional exponent depends on the coordinates and time. Therefore, we
have to introduce integral functionals (both left-sided and right-sided)
which are suitable to describe the dynamics of functions defined on
multifractal sets (see [1]). Actually, this functionals are simple and
natural generalization the Riemann-Liouville fractional derivatives and
integrals:
\begin{equation} \label{1}
D_{+,t}^{d}f(t)=\left( \frac{d}{dt}\right)^{n}\int_{a}^{t}
\frac{f(t^{\prime})dt^{\prime}}{\Gamma
(n-d(t^{\prime}))(t-t^{\prime})^{d(t^{\prime})-n+1}}
\end{equation}
\begin{equation} \label{2}
D_{-,t}^{d}f(t)=(-1)^{n}\left( \frac{d}{dt}\right)
^{n}\int_{t}^{b}\frac{f(t^{\prime})dt^{\prime}}{\Gamma
(n-d(t^{\prime}))(t^{\prime}-t)^{d(t^{\prime})-n+1}}
\end{equation}
where $\Gamma(x)$ is Euler's gamma function, and $a$ and $b$ are
some constants from $[0,\infty) $. In these definitions, as
usually, $n=\{d\}+1$ , where $\{d\}$ is the integer part of $d$ if
$d\geq 0$ (i.e. $n-1\le d<n$) and $n=0$ for $d<0$. If $d=const$,
the generalized fractional derivatives (GFD) (\ref{1})-(\ref{2})
coincide with the Riemann - Liouville fractional derivatives
($d\geq 0$) or fractional integrals ($d<0$). When $d=n+\varepsilon
(t),\, \varepsilon (t)\rightarrow 0$, GFD can be represented by
means of integer derivatives and integrals. For $n=1$, that is,
$d=1+\varepsilon$, $\left| \varepsilon \right| <<1$ it is possible
to obtain:
\begin{equation} \label{3}
D_{+,t}^{1+\varepsilon }f(t)\approx \frac{\partial}{\partial t}
f(t)+a\frac{\partial}{\partial t}\left[\varepsilon (r(t),t)f(t)\right]
\end{equation}
where $a$ is constant and defined by the choice of the rules of
regularization of integrals (\ref{1})-(\ref{2}) (for more detailed see
\cite{kob1}). The selection of the rule of regularization that gives a
real additives for usual derivative in (\ref{3}) yeilds $a=0.5$ for $d<1$
and $a=1.077$ for $d>1$ \cite{kob1}. The functions under integral sign in
(\ref{1})-(\ref{2}) we consider as the generalized functions defined on
the set of the finite functions \cite{gel}. The notions of GFD, similar to
(\ref{1})-(\ref{2}), can also be defined for the space variables ${\mathbf
r}$.

The definitions of GFD (\ref{1})-(\ref{2}) are formal until the
connections between fractal dimensions of time $d_{t}({\mathbf r}(t),t)$
and certain characteristics of physical fields (say, potentials $\Phi
_{i}({\mathbf r}(t),t),\,i=1,2,..)$ or densities of Lagrangians $L_{i}$)
are determined. Following \cite{kob1}, we define this connection by the
relation
\begin{equation} \label{4}
d_{t}({\mathbf r}(t),t)=1+\sum_{i}\beta_{i}L_{i}(\Phi_{i}
({\mathbf r}(t),t))
\end{equation}
where $L_{i}$ are densities of energy of physical fields, $\beta_{i}$ are
dimensional constants with physical dimension of $[L_{i}]^{-1}$ (it is
worth to choose $\beta _{i}^{\prime}$ in the form $\beta _{i}^{\prime
}=a^{-1}\beta _{i}$ for the sake of independence from regularization
constant). The definition of time as the system of subsets and definition
(\ref{4}) put the value of fractional (fractal) dimensionality
$d_{t}(r(t),t)$ into accordance with every time instant $t$. The latter
depends both on time $t$ and coordinates ${\mathbf r}$. If $d_{t}=1$
(absence of physical fields) the set of time has topological
dimensionality equal to unity. The multifractal model of time allows, as
will be shown below, to consider the divergence of energy of masses moving
with speed of light in the SR theory, as the result of the requirement of
rigorous validity, rather than approximate fulfillment, of the laws
pointed out in the beginning of this paper in the presence of physical
fields.

\section{The principle of the velocity of light invariance}

Because of the inhomogeneity of time in our multifractal model, the speed
of light, just as in the general relativity theory, depends on potentials
of physical fields that define the fractal dimensionality of time
$d_{t}({\mathbf r}(t),t)$ (see (\ref{4})). If fractal dimensionality
$d_{t}({\mathbf r}(t),t)$ is close enough to unity
($d_{t}(r(t),t)=1+\varepsilon, \, \left| \varepsilon \right|<<1$), the
difference of the speed of light in moving (with velocity $v$ along the
$x$ axis) and fixed frame of reference will be small. In the systems that
move with respect to each other with almost constant velocity (stationary
velocities do not exist in the mathematical theory based on definitions of
GFD (\ref{1}) - (\ref{2})) the speed of light can not be taken as a
fundamental constant. In the multifractal time theory the principle of the
speed of light invariance can be considered only as approximate. But if
$\varepsilon$ is small, it allows to consider a nonlinear coordinates
transformations from the fixed frame to the moving (replacing the
transformations of Galilee in inhomogeneous time and space), as close to
linear (weakly nonlinear) transformations and, thus, makes it possible to
preserve the conservation laws, and all the invariants of the Minkowski
space, as the approximate laws. Then the way of reasoning and
argumentation accepted in SR theory (see for example, \cite{mat}) can also
remain valid. Designating the coordinates in the moving and fixed frames
of reference through $x^{\prime}$ and $x$, accordingly, we write down
\begin{eqnarray}  \label{5} \nonumber
x^{\prime}&=&\alpha (t,x)[x-v(x,t)t(x(t),t] \\
x&=&\alpha^{\prime}(t,x)[x^{\prime}+
v^{\prime}(x^{\prime}(t^{\prime}),t^{\prime}),
\,t^{\prime}(x^{\prime}(t^{\prime}),t^{\prime})
\end{eqnarray}
In (\ref{5}) $\alpha \neq \alpha^{\prime}$ and the velocities $v^{\prime}$
and $v$ (as well as $t$ and $t^{\prime}$) are not equal (it follows from
the inhomogenity of multifractal time). Place clocks in origins of both
the frames of reference and let the light signal be emitted in the moment,
when the origins of the fixed and moving frames coincide in space and time
at the instant $t^{\prime}=t=0$ and in points $x^{\prime}=x=0$. The
propagation of light in moving and fixed frames of reference is then
determined by equations
\begin{equation} \label{6}
x^{\prime}=c^{\prime}t^{\prime}\,\,\,x=ct
\end{equation}
These characterize the propagation of light in both of the frames of
reference at every moment. Due to the time inhomogenity $c^{\prime} \neq
c$, but since $\left| \varepsilon <<1 \right|$ the difference between
velocities of light in the two frames of reference will be small. For this
case we can neglect the distinction between $\alpha^{\prime}$ and $\alpha
$ and, for different frames of reference  write the expressions for
velocities of light, using (\ref{3}) to define velocity (denote $f(t)=x,\,
dx/dt=c_{0}$). Thus we obtain
\begin{equation} \label{7}
c=D_{+,t}^{1+\varepsilon}x=c_{0}(1-\varepsilon)-
\frac{d\varepsilon}{dt}x
\end{equation}
\begin{equation} \label{8}
c^{\prime}=D_{+,t}^{1+\varepsilon^{\prime}}x^{\prime}=
c_{0}(1-\varepsilon)+ \frac{d\varepsilon}{dt}x^{\prime}
\end{equation}
\begin{equation} \label{9}
c_{1}= c_{0}(1-\varepsilon)- \frac{d\varepsilon}{dt}x^{\prime}
\end{equation}
\begin{equation} \label{10}
c_{1}^{\prime}= c_{0}(1-\varepsilon)+ \frac{d\varepsilon}{dt}x
\end{equation}
The equalities (\ref{9}) and (\ref{10}) appear in our model of
multifractal time as the result of the fact, that in this model all the
frames of reference are absolute frames of reference (because of material
character of the time field) and the speed of light depends on the state
of frames: if the frame of reference is a moving or a fixed one, if the
object under consideration in this frame moves or not. This dependence
disappears only when $\varepsilon=0$. Before substitution the relations
(\ref{5}) in the equalities (\ref{7}) - (\ref{10}) (with
$\alpha^{\prime}\approx\alpha$) it is necessary to find out how
$d\varepsilon/dt$ depends on $\alpha$. Using for this purpose Eq.(\ref{4})
we obtain:
\begin{equation} \label{11}
\frac{d\varepsilon}{dt}=\frac{d\varepsilon}{d{\mathbf r}} {\mathbf
v} \approx -\sum_{i}\beta_{i}({\mathbf F}_{i} {\mathbf v}
+\frac{\partial L_{i}}{\partial t})
\end{equation}
where ${\mathbf F }_{i}=dL_{i}/d{\mathbf r}$. Since the forces for moving
frames of reference are proportional to $\alpha$ we get (for the case when
there is no explicit dependence of $L_{i}$ on time)
\begin{equation}   \label{12}
\frac{d\varepsilon}{dt}\approx -\sum_{i}\beta_{i}{\mathbf F}_{0i}
{\mathbf v} \alpha
\end{equation}
where $F_{0i}$ are the corresponding forces at zero velocity. Multiplying
(\ref{7}) - (\ref{10}) on the corresponding times
$t,\,t^{\prime},\,t_{1},\,t_{1}^{\prime}$ yields the following expressions
\begin{equation} \label{13}
c^{\prime}t^{\prime}=c_{0}t\left[1+\frac{v\sum_{i}\beta_{i}F_{0i}}
{c_{0}}\alpha^{2}ct(1-\frac{v}{c})\right]
\end{equation}
\begin{equation} \label{14}
ct=c_{0}t^{\prime}\left[1+\frac{v\sum_{i}\beta_{i}F_{0i}}
{c_{0}}\alpha^{2}ct(1+\frac{v}{c})\right]
\end{equation}
\begin{equation} \label{15}
c_{1}^{\prime}t_{1}^{\prime}=c_{0}t_{1}\left[1-\frac{v\sum_{i}\beta_{i}F_{0i}}
{c_{0}}\alpha^{2}ct(1-\frac{v}{c})\right]
\end{equation}
\begin{equation} \label{16}
c_{1}t_{1}=c_{0}t_{1}^{\prime}\left[1-\frac{v\sum_{i}\beta_{i}F_{0i}}
{c_{0}}\alpha^{2}ct(1+\frac{v}{c})\right]
\end{equation}
Since in our model the motion and frames of reference are
absolute, the times $t_{1}$ and $t_{1}^{\prime}$ correspond to the
cases, when the moving and fixed frames of reference exchange
their roles - the moving one becomes fixed and vice versa. These
times coincide only when $\varepsilon=0$. The times in square
brackets, as well as the velocities, are taken to equal, because
the terms containing them are small as compared to unity. The
principle of invariance of the velocity of light for transition
between the moving and fixed frames of reference in multifractal
time model is approximate (though quite natural, because the
frames of reference are absolute frames of reference). Taking into
account (\ref{5}), the relations (\ref{13}) - (\ref{16}) take the
form
\begin{equation} \label{17}
c^{\prime}t^{\prime}=c\alpha t(1-\frac{v}{c}),\,\,\,\,
c_{1}^{\prime}t_{1}^{\prime}=c\alpha t_{1}(1-\frac{v}{c})
\end{equation}
\begin{equation} \label{18}
ct=c\alpha t^{\prime}(1+\frac{v}{c}),\,\,\,\, c_{1}t_{1}=c\alpha
t^{\prime}(1+\frac{v}{c})
\end{equation}
Once again we note, that the four equations for
$c_{1}^{\prime}t_{1}^{\prime}$ and $c_{1}t_{1}$, instead of the two
equations in special relativity, appear as the consequence of the absolute
character of the motion and frames of reference in the model of
multifractal time. In the right-hand side of (\ref{17}) - (\ref{18}) the
dependence of velocity of light on fractal dimensions of time is not taken
into account (just as in the equations (\ref{13}) - (\ref{16})). Actually,
this dependence leads to pretty unwieldy expressions. But if we retain
only the terms that depend on $\beta=\sqrt{|1-v^{2}/c^{2}|}$ or $a_{0}$
and neglect unessential terms containing the products $\beta \alpha_{0}$,
utilizing (\ref{13}) - (\ref{16}) after the multiplication of the four
equalities (\ref{17}) - (\ref{18}), we receive the following equation for
$\alpha$ (it satisfies to all four equations):
\begin{equation} \label{19}
4a_{0}^{4}\beta^{4}\alpha^{8}-4a_{0}^{2}\alpha^{4}+1=
\beta^{4}\alpha^{4}+4a_{0}^{4}\beta^{4}\alpha^{8}
\end{equation}
where
\begin{equation} \label{20}
\beta=\sqrt{\left|1-\frac{v^{2}}{c^{2}} \right|}
\end{equation}
\begin{equation} \label{21}
a_{0}=\sum_{i}\beta_{i}F_{0i}\frac{v}{c}ct
\end{equation}
From (\ref{19}) follows
\begin{equation} \label{22}
\alpha_{1}\equiv\beta^{*^{-1}}= \frac{1}{\sqrt[4]{\beta^{4}+4a_{0}^2}}
\end{equation}
The solutions $\alpha_{2,3,4}$ are given by
$\alpha_{2}=-\alpha_{1},\,\alpha_{3,4}=\pm i\alpha$. Applicability of
above obtained results is restricted by requirement $|\varepsilon| \ll 1$

\section{Lorentz transformations and transformations of length \\
and time in multifractal time model}

The Lorentz transformations, as well as transformations of coordinate
frames of reference, in the multifractal model of time are nonlinear due
to the dependence of the fractional dimensions of time $d_{t}({\mathbf
r},t)$ on coordinates and time. Since the nonlinear corrections to Lorentz
transformation rules are very small for $\varepsilon \ll 1$, we shall take
into account only the corrections that eliminate the singularity at the
velocity $v=c$. It yields in the replacement of the factor $\beta^{-1}$ in
Lorentz transformations by the modified factor $\alpha=1/\beta^{*} $ given
by (\ref{22}). The Lorentz transformation rules (for the motion along the
$x$ axis) take the form
\begin{equation}     \label{23}
x^{\prime}=\frac{1}{\beta^{*}}(x-vt),\,\,\,\,\,
t^{\prime}=\frac{1}{\beta^{*}}(t-x\frac{v}{c^{2}})
\end{equation}
In the equations (\ref{22}) and (\ref{23}) the velocities $v$ and $c$
weakly depend on $x$ and $t$  and their contribution to the singular terms
is small. Hence, we can neglect this dependence. The transformations from
fixed system to moving system are almost orthogonal (for $\varepsilon \ll
1$ ), and the squares of almost four-dimensional vectors of Minkowski
space vary under the coordinates transformations very slightly (i.e. they
are almost invariant). Then it is possible to neglect the correction terms
of order about $O(\varepsilon,\dot{\varepsilon})$, which, for not equal to
infinity variables, are very small too. From (\ref{22}) - (\ref{23}) the
possibility of arbitrary velocity motion of bodies with nonzero rest mass
follows. With the corrections of order $O(\varepsilon,\dot{\varepsilon})$
in nonsingular terms being neglected, the momentum and energy of a body
with a nonzero rest mass in the frame of reference moving along the $x$
axis ($E_{0}=m_{0}c^{2})$ equal to
\begin{equation} \label{24}
p=\frac{1}{\beta^{*}}m_{0}v=\frac{m_{0}v}{\sqrt[4]{\beta^{4}+4a_{0}^{2}}},\,\,\,
E=E_{0}\sqrt{\frac{v^{2}c^{-2}}{\sqrt{\beta^{4}+4a_{0}^{2}}}+1}
\end{equation}
The energy of such a body reaches its maximal value at $v=c$ and is equal
then $E_{v=c}\approx E_{0}/\sqrt{2\alpha_{0}}$. When $v \to \infty$ the
energy is finite an tends to $E_{0}\sqrt{2}$. For $v \le c$ the total
energy of a body is represented by the expression
\begin{equation} \label{25}
E\cong\frac{E_{0}}{\sqrt[4]{\beta^{4}+4 a_{0}^{2}}}=mc^{2},\,\,\,
m=\frac{m_{0}}{\beta^{\ast}}
\end{equation}
For $v \ge c$, total energy, defined by (\ref{24}), is given by
\begin{eqnarray}  \label{26} \nonumber
E&=&mc^{2},\,\,\,\, \beta^{2}=\frac{v^{2}}{c^{2}}-1  \\ m&=&
\frac{m_{0}}{\beta^{\ast}}\sqrt{2v^{2}/c^{2}+4a_{0}^2}
\end{eqnarray}
If we are to take into account only the gravitational field of Earth
(here, as in \cite{log}, gravitational field is a real field) and neglect
the influence of all the other fields), the parameter $a_{0}(t)$ can be
estimated to be $a_{0}=r_{0}r^{-3}x_{E}c t$, where $r_{0}$ is the
gravitational radius of Earth, $r$ is the distance from the Earth's
surface to its center ($\varepsilon=0.5\beta_{g}\Phi_{g},\,
\beta_{g}=2c^{-2},\,x_{E}\sim r,\,v=c$). For energy maximum we get
$E_{max}\sim E_{0}\cdot10^{3}t^{-0.5}sec^{0.5}$.

Shortening of lengths and time intervals in moving frames of reference in
the model of multifractal time also have several peculiarities. Let $l$
and $t$ be the length and time interval in a fixed frame of reference. In
a moving frame
\begin{equation} \label{27}
l^{\prime}=\beta^{\ast} l,\,\,\, t^{\prime}=\beta^{\ast}t
\end{equation}
Thus, there exist the maximal shortening of length when the body's
velocity equals to  the speed of light. With the further increasing of
velocity (if it is possible to fulfill some requirements for a motion in
this region with constant velocity without radiating), the length of a
body begins to grow and at infinitely large velocity is also infinite. The
slowing-down of time, from the point of view of the observer in the fixed
frame (maximal shortening equals to $t^{\prime}=t\sqrt{2a_{0}}$) is
replaced, with the further increase of velocity beyond the speed of light,
by acceleration of time passing ($t \to 0$ when $v \to \infty$).

The rule for velocities transformation retains its form, but
$\beta$ is replaced by $\beta^{*}$
\begin{equation} \label{28}
u_{x}=\frac{u_{x}^{\prime}+v}{1+\frac{u_{x}^{\prime} v}{c^{2}}},\,
u_{y}=\frac{u_{y}^{\prime}\beta^{\ast}}{1+\frac{u_{y}^{\prime}
v}{c^{2}}},\,
u_{z}=\frac{u_{z}^{\prime}\beta^{\ast}}{1+\frac{u_{z}^{\prime} v}{c^{2}}}
\end{equation}
Since there is no law that prohibits velocities greater than that
of light, the velocities in (\ref{28}) can also exceed the speed
of light. The electrodynamics of moving media in the model of
multifractal time can be obtained, in most cases, by the
substitution $\beta\to\beta^{*}$.

\section{Conclusions}

To conclude, the theory of relative motions in almost inertial
systems based on the multifractal time theory \cite{kob1} is
invented. This theory describes open systems (for statistical
theory of open systems see in \cite{klim} ) and in this theory
motion with any velocity is possible. The theory coincides with
special relativity after transition to inertial systems (if we
neglect the fractional dimensions of time) or almost coincides
(the differences are negligible) for velocities $v<c$. Movement of
bodies with velocities that exceed the speed of light is
accompanied by a number of physical effect's which can be found
experimentally (these effects will be considered in the separate
paper in more detail).

\end{document}